\documentclass{aastex}

\newcommand{\kms}{km~s$^{-1}$}

\newcommand{\lya}{Ly$\alpha$}

\newcommand{\zem}{$z_{em}$}
\newcommand{\zabs}{$z_{abs}$}
\newcommand{\civ}{{\rm C}~{\sc iv}}

\newcommand{\cii}{{\rm C}~{\sc ii}}
\newcommand{\oi}{{\rm O}~{\sc i}}

\newcommand{\alii}{{\rm Al}~{\sc ii}}

\newcommand{\siii}{{\rm Si}~{\sc ii}}
\newcommand{\feii}{{\rm Fe}~{\sc ii}}

\slugcomment{Accepted for publication in the Astronomical Journal (AJ)}
\shorttitle{Subaru high resolution spectroscopy of HS1603+3820}
\shortauthors{Misawa et al.}

\begin{document}

\title{Subaru high resolution spectroscopy of complex metal absorption
lines of QSO HS1603+3820\footnotemark[1]}

\footnotetext[1]{Based on data collected at Subaru Telescope, which is
operated by the National Astronomical Observatory of Japan.}

\author{Toru Misawa\altaffilmark{2,3}, 
        Toru Yamada\altaffilmark{4}, 
        Masahide Takada-Hidai\altaffilmark{5},
        Yiping Wang\altaffilmark{6},
        Nobunari Kashikawa\altaffilmark{4},
        Masanori Iye\altaffilmark{4,7}, and
        Ichi Tanaka\altaffilmark{4}
}

\altaffiltext{2}{Department of Astronomy, School of Science,
 University of Tokyo, 7-3-1 Hongo, Bunkyo-ku, Tokyo 113-0033, Japan}
\altaffiltext{3}{misawatr@cc.nao.ac.jp}
\altaffiltext{4}{National Astronomical Observatory, 2-21-1 Osawa,
 Mitaka, Tokyo 181-8588, Japan}
\altaffiltext{5}{Liberal Arts Education Center, Tokai University, 1117
 Kitakaname, Hiratsuka, Kanagawa 259-1292, Japan}
\altaffiltext{6}{Purple Mountain Observatory, National Astronomical
 Observatories, China}
\altaffiltext{7}{Department of Astronomical Science, The Graduate
 University for Advanced Stuides, 2-21-1 Osawa, Mitaka, Tokyo
 181-8588, Japan.}

\begin{abstract}
 We present a high resolution spectrum of the quasar, HS1603+3820
($z_{em}$=2.542), observed with the High Dispersion Spectrograph (HDS)
on Subaru Telescope. This quasar, first discovered in the Hamburg/CfA
Quasar Survey, has 11 \civ\ lines at 1.96 $<$ \zabs $<$ 2.55. Our
spectrum covers 8 of the 11 \civ\ lines at \zabs\ $>$ 2.29 and
resolves some of them into multiple narrow components with b $<$ 25
\kms\ because of the high spectral resolution R=45000, while other
lines show broad profiles (b $>$ 65 \kms). We use three properties of
\civ\ lines, specifically, time variability, covering factor, and
absorption line profile, to classify them into quasar intrinsic
absorption lines (QIALs) and spatially intervening absorption lines
(SIALs). The \civ\ lines at 2.42 $<$ \zabs $<$ 2.45 are classified as
QIALs in spite of their large velocity shifts from the quasar. Perhaps
they are produced by gas clouds ejected from the quasar with the
velocity of $v_{ej}=$ 8000 \kms $-$ 10000 \kms. On the other hand,
three \civ\ lines at 2.48 $<$ \zabs $<$ 2.55 are classified as SIALs,
which suggests there exist intervening absorbers near the quasar. We,
however, cannot rule out QIALs for the two lines at \zabs\ $\sim$ 2.54
and 2.55, because their velocity shifts, 430 \kms\ blueward and 950
\kms\ redward of the quasar, are very small. The \civ\ line at \zabs\
$\sim$ 2.48 consists of many narrow components, and has also
corresponding low-ionization metal lines (\alii, \siii, and
\feii). The velocity distribution of these low-ionization ions is
concentrated at the center of the system compared to that of
high-ionization \civ\ ion. Therefore we ascribe this system of
absorption lines to an intervening galaxy.

\end{abstract}


\section{Introduction}
The bright quasar, HS1603+3820 ($z_{em}$=2.542, B=15.9), which was
first discovered in the Hamburg/CfA Bright Quasar Survey (Hagen et
al. 1995; Dobrzycki et al. 1996), has very unique properties, namely
the combination of high redshift, large luminosity, and richness of
metal absorption lines. In the spectrum of this quasar, 11 \civ\
absorption lines were detected at 1.965 $<$ \zabs $<$ 2.554
(Dobrzycki, Engels, \& Hagen 1999; hereafter D99). Among the 11 \civ\
lines, 8 lines are observed at 2.420 $<$ \zabs $<$ 2.554, which
corresponds to a velocity shift of $-$10600 \kms $< \Delta v <$
$+$1000 \kms\ from the quasar. The number density of \civ\ lines with
the rest-frame equivalent width of $W_{rest} \geq 0.15$ is $N(z) \sim
21$ per unit redshift at \zabs\ $\sim$ 2.38 with $\Delta z$ = 0.34. If
we combine \civ\ lines that lie within 1000 \kms\ of each other to
produce a so-called Poisson sample (cf. Sargent, Boksenberg, \&
Steidel 1988), $N(z)$ becomes $\sim$ 12, which is significantly larger
than the expected average value, $N(z)=2.45^{+0.60}_{-0.49}$, at
\zabs\ $\sim$ 2.40 evaluated for \civ\ lines with $W_{rest} \geq\
0.15$ \AA\ and $v_{ej} > 5000$ \kms\ in the previous studies (e.g.,
Sargent et al. 1988; Steidel 1990; Misawa et al. 2002). Such an
extreme overdensity of \civ\ lines inspired us to get a high
resolution spectrum of this quasar with Subaru Telescope for a
detailed study of the \civ\ lines.

Absorbers are generally divided into quasar intrinsic absorption lines
(QIALs) and spatially intervening absorption lines (SIALs). QIALs are
thought to be produced by gas clouds that are intrinsically associated
with the nuclear regions of the quasars. On the other hand, SIALs are
produced by either intervening galaxies, which are located along the
line of sight of the quasars, or the interstellar matter in the quasar
host galaxies. It is essential to understand the origins of the
absorption lines in order to answer the questions, such as what causes
the extremely high number density of \civ\ lines in the spectrum of
HS1603+3820.

To distinguish QIALs from SIALs, we use the seven proper criteria
which have been proposed in the literatures (Barlow \& Sargent 1997;
Hamann et al. 1997a; and references therein):

(1) Time variability of equivalent width is sometimes detected in
    QIALs.

(2) QIALs are produced by absorbers with high electron space density.

(3) Absorbers that produce QIALs cover the background the quasar
    partially.

(4) Polarization rate in QIALs is larger than that in the continuum.

(5) QIALs have smoother and broader line profiles than those of SIALs.

(6) QIALs are produced by high ionized absorbers.

(7) QIALs are produced by absorbers with high metallicity.

With one or combination of these criteria, for example, Barlow et
al. (1992) found that the broad absorption line (BAL) with FWHM $\geq$
100 \kms\ near the emission redshift of CSO203 is a QIAL, because the
line shows time variability. Hamann, Barlow, \& Junkkarinen (1997b)
also found that the broad \civ\ doublet even at $-$24000 \kms\ from
Q2343+125 show time variability. Barlow \& Sargent (1997) and Hamann
et al. (1997a) found QIALs by the covering factor analysis in
PKS0123+257 and UM675. Goodrich \& Miller (1995) used the polarization
method to distinguish QIALs from SIALs in PHL5200 and H1413+117.

For  HS1603+3820, D99 used only the covering factor analysis with
their low resolution spectrum. They applied the method to the broad
\civ\ line at the high ejection velocity end of the \civ\ complex, and
found that it is a QIAL. However, the other lines were not classified,
because they have line profiles narrower than the spectral resolution
or maybe they are affected by other lines.

In this paper, we present a high resolution (R $\sim$ 45000) spectrum
of HS1603+3820 obtained with the High Dispersion Spectrograph (HDS;
Noguchi et al. 2002) on the Subaru Telescope. The \civ\ lines are
found to have further complex structure, and we classify them into
QIALs and SIALs in order to discuss the cause of their large number
density.  In \S\ 2, we describe the observations and data
reduction. In \S\ 3, the properties of \civ\ systems are examined, and
the classification of \civ\ systems is delineated in \S\ 4. Results
and discussion are given in \S\ 5. We present the conclusions and the
prospect of the future work in \S\ 6.

\section{Observation and Data reduction}

We carried out spectroscopic observation of the quasar HS1603+3820
($\alpha_{2000}=$16h04m55$^{\prime}$4, $\delta_{2000}=
+38^{\circ}12^{\prime}01^{\prime \prime}$) with HDS on Subaru
Telescope on 2002 March 23 (UT).

It has to be noted that the \lya\ emission line is affected by complex
metal absorption lines, which makes the value of emission-line
redshift of the quasar uncertain (D99). D99 suggested the emission
redshift as \zem\ $\sim$ 2.51 by estimating a continuum around the
\lya\ and \civ\ emission lines, and considering the fact that the
redshifts determined from the broad emission lines can be blueshifted
by $\geq$ 1200 \kms\ with respect to the systemic redshift from the
narrow forbidden lines (Espey 1993). If we match the spectrum of the
quasar taken by D99 to the SDSS composite spectrum (Vanden Berk et
al. 2001) using \oi\ $\lambda 1304$ $+$ \siii\ $\lambda 1307$ and
\cii\ $\lambda 1335$ emission lines, the systemic redshift of the
quasar is estimated to be \zem\ = 2.542 $\pm$ 0.003. Richards et
al. (2002a) also showed that quasars with weaker \civ\ emission lines
tend to be large blueshifted in excess of 2000 \kms. Since the \civ\
emission line of HS1603+3820 is very weak, the systemic redshift of
the quasar may be larger than \zem\ = 2.51. Therefore we assume \zem\
of the quasar to be 2.542 instead of \zem\ = 2.51 throughout this
paper.

HDS has two 2k $\times$ 4k CCDs which are the blue and red CCDs. We
adopted 2 pixel binning along the slit, and we used the red grating
with a central wavelength of 6450 \AA, which covers 5100 \AA\ -- 6400
\AA\ on the blue CCD and 6500 \AA\ -- 7600 \AA\ on the red CCD. The
results discussed in this paper are based on the spectra taken by the
blue CCD alone. The slit width of 0$^{\prime\prime}$.8 provides the
spectral resolution of 45000 (6.67 \kms). The weather conditions were
very good, and the seeing was $\sim$ 0$^{\prime\prime}$.5. The S/N
ratio of the spectrum with 2700 sec exposure was $\sim$ 40 per pixel
around $\lambda$ = 5450 \AA.

The data were reduced with the IRAF package in the standard manner. We
determined the continuum by fitting the data with a 3rd order cubic
spline function. The spectrum has a heavily absorbed region at around
5300 \AA\ -- 5350 \AA\ in the 20th echelle order (Figure 1) and it is
not straightforward to determine the continuum level there. In such
case, we evaluated the continuum profile of the heavily absorbed order
$i$, $q_{i}(m)$, by
\begin{equation}
 q_{i}(m) = \frac{\frac{f_{i}(m)}{f_{i-1}(m)} \times q_{i-1}(m) + 
                  \frac{f_{i}(m)}{f_{i+1}(m)} \times q_{i+1}(m)}{2},
\label{eqn:1}
\end{equation}
where $f_{i-1}(m)$, $f_{i}(m)$, and $f_{i+1}(m)$ are the counts at the
$m$-th pixel of order $i-1$, $i$ and $i+1$ of the continuum profile
fitted for the flat-frame spectrum, $q_{i-1}(m)$, and $q_{i+1}(m)$ are
the counts of order $i-1$ and $i+1$ of the continuum profile fitted
for the quasar spectrum.  We acquired the normalized $i$-th order
spectrum by dividing the quasar spectrum of order $i$ by $q_{i}(m)$.
We used order 17 and 23 to produce $q_{20}(m)$, because orders 18, 19,
21, 22 are weakly affected by absorption lines. In the middle window
of Figure 1, the interpolated continuum of the order 20, $q_{20}(m)$,
is shown as a dotted line.
For confirmation of the validity of this technique, we applied this
method to the stellar spectrum that was used to locate the positions
and track the shapes of all the echelle orders. The continuum flux of
order $i$ obtained with this interpolation method is only 3.3 \%
smaller on average than that of directly fitted spectrum. Therefore,
we believe this technique is sufficiently useful for our study.

The normalized spectrum of each order is combined to produce the final
spectrum (Figure 2).

\section{Identification of \civ\ systems}
We used the VPFIT Voigt profile line-fitting software developed by
Carswell et al. (Webb 1987; Carswell et al. 1987). D99 detected 11
\civ\ lines at 1.96 $<$ \zabs\ $<$ 2.55. Our spectrum covers 8 of them
in the range at $2.42 < z_{abs} < 2.55$ and detected absorption lines
at the wavelength of all these lines.  Thanks to high spectral
resolution, some of \civ\ lines detected in D99 are further resolved
into multiple narrow components, resulting in a total 30 \civ\
components between $z$ = 2.42 and 2.55. Figure 3 shows both the HDS
and MMT spectra (D99; taken from the web site,
http://hea-www.harvard.edu/QEDT/Papers/hs1603). We summarize the
results in Table 1. Column (1) and (2) are the metal line
identification and ID number. Column (3) and (4) are observed
wavelength and redshift. Column (5) gives the velocity difference from
$z_{em}$, and column (6) is the Doppler parameter, $b = \sqrt{2}
\sigma$ ($\sigma$ is velocity dispersion). Columns (7) to (10) are
covering factor, optical depth at line center, and column densities in
the case of the covering factor $C_{f}=1$ and $C_{f} < 1$, which are
discussed in \S\ 4 in detail. 

For the sake of simplicity, we divided these \civ\ lines into four
systems, System A to System D as shown in Figure 3. The \civ\ lines in
System B and D in D99 are resolved into multiple narrow components
with $b < 25$ \kms\ in our spectrum. The line in System C is not
resolved but this line  has small Doppler value $b = 11$ \kms. On the
other hand, the five \civ\ line clustering in System A in D99 are not
further resolved into narrow components, despite that three of them
have broad line profiles, $b > 70$ \kms. Detailed note on each system
is described below.

System A (2.42 $< z_{abs} <$ 2.45): D99 detected 5 \civ\ lines in this
system at the velocity shift of $-$10600 \kms\ $< \Delta v <$ $-$7800
\kms\ from $z_{em}$ in their spectrum. We found additional two \civ\
lines in our spectrum as shown in Figure 4.  In addition to these
lines, there still remains the broad absorption feature at $-$9300
\kms\ $< \Delta v <$ $-$8000 \kms. We cannot separate them to fit the
Voigt profile individually because they are heavily blended with each
other. Five of the seven detected \civ\ lines (\zabs\ = 2.4186,
2.4257, 2.4341, 2.4362, and 2.4510) are relatively broad and smooth,
and their appearance are quite similar to each other. On the other
hand, the line (\zabs\ = 2.4366) has relatively narrow line
profile. Perhaps it is an independent system of the broad \civ\
lines. Another interesting result is that \feii\ $\lambda 1608$ at
$z_{abs} = 2.4367$ detected in D99 have disappeared in our
spectrum. But this line could be just a glitch because it looks
narrower than the resolution of the spectrum in D99.

System B ($z_{abs}$=2.48): This is the strongest \civ\ line in the MMT
spectrum (D99). Our spectrum resolve this line into 18 narrow
components with relatively small Doppler parameters of 4 $\leq b \leq$
25 \kms\ (Figure 5), and most of them can be explained by thermal
broadening. Relatively low-ionization lines such as \siii\ $\lambda
1526$, \alii\ $\lambda 1670$, and, \feii\ $\lambda 1608$ associated
with the \civ\ lines are also detected (Figure 5). The strongest
components of the low-ionization lines are observed at the center of
this system, while the weak components exist almost symmetrically at
both sides of the strongest one. 

System C ($z_{abs}$=2.54): This is one of the two systems within 1000
\kms\ from the quasar. The line itself is narrow and not resolved into
multiple components (Figure 6). There are no other significant metal
lines at this redshift.

System D ($z_{abs}$=2.55): Another system within 1000 \kms\ from the
quasar, though this system is redshifted from the quasar. This system
is resolved into 4 narrow components (Figure 7). The central two
components reach almost zero intensity at their bottoms. The system
also do not show up in any other metal lines.

\section{Classification of \civ\ lines}

We detected as many as 30 \civ, 7 \alii, 8 \siii, and one \feii\ lines
with velocity difference in the range of $-10600$ \kms\ $< \Delta v <
+1000$ \kms\ from the quasar.  While all the metal lines discovered
are, in terms of the velocity difference, relatively close to the
quasar, they could be produced by either gas clouds that are
intrinsically associated with the quasar or in intervening
galaxies. With the HDS spectrum, we examine the line profiles in
detail to classify them into QIAL and SIAL. Among the seven criteria
to distinguish QIALs from SIALs described in \S 1, we  mainly use the
criteria (1), (3) and (5) which can be directly applied to our
spectrum.

\subsection{Absorption line profile}

Absorbers which produce very broad and smooth profiles are expected to
be intrinsically associated with the quasar nuclear region. For
example, the broad absorption line systems (BALs) with $b \sim\
10^{4}$ \kms\ are considered to be originated in gas motion in broad
line region (BLR) of quasars. On the other hand, intervening
absorption lines that inherent to cosmologically intervening galaxies
have narrow profile with $b \leq 12$ \kms, which corresponds to gas
temperature, T $\sim 10^{5}$ K for \civ\ ion.

System A has five relatively broad \civ\ lines ($b \geq 65$ \kms), and
all of them have relatively smooth line profiles and are not resolved
into narrow components even in our high-resolution spectrum. Their
broad line profiles with $b$ ($=\sqrt{b_{T}^{2}+b_{tur}^{2}}$) $\geq
65$ \kms\ are not likely to be produced by only thermal broadening,
$b_{T}$, because the corresponding temperature is very high, T $\geq 3
\times\ 10^{6}$ K, compared with typical metal line systems. The micro
turbulence, $b_{tur}$, is expected to contribute to the majority of
the Doppler parameter. Therefore the results strongly suggest that
these lines are intrinsic ones which occur in the vicinity of the
quasar nucleus although Doppler parameters of them are too small
compared with those of BALs.

On the other hand, System B, C, and D consist of only narrow
components with $b \leq 25$ \kms\ (corresponding temperature is T
$\leq 4.5 \times 10^{5}$ K for \civ\ ion) and 65$\%$ of them have $b <
12$ \kms, which imply that these systems can be produced by
intervening galaxies. The component at \zabs\ = 2.4412 in System A
also have relatively narrow line profile.

\subsection{Partial coverage}

We evaluate the covering factor, $C_{f}$, which is the line-of-sight
coverage fraction of the absorber over the continuum source. Since the
components of absorption systems seem to be fully resolved in our
spectrum, we can evaluate the covering factors for \civ\ doublets,
using the following equations:
\begin{equation}
C_{f} = \left\{
\begin{array}{lll}
 \frac{R_{r}^{2}-2R_{r}+1}{R_{b}-2R_{r}+1} & for & R_{r} > R_{b} \geq R_{r}^{2} \\
 1                                         & for & R_{b} < R_{r}^{2} \\
 1 - R_{r}                                 & for & R_{b} \geq R_{r}
\end{array}
\label{eqn:2}
\right.
\end{equation}
where $R_{b}$ and $R_{r}$ are residual fluxes at the line centers for
blue (\civ\ $\lambda 1548$) and red (\civ\ $\lambda 1551$)  components
of the doublet lines on the normalized spectrum (Hamann et al. 1997a;
Barlow \& Sargent 1997).

We also evaluated optical depth at line center, $\tau$, and the total
column density, $\log N$, for blue component derived by the following
equations (Hamann et al. 1997a; Barlow \& Sargent 1997; Savage \&
Sembach 1991; Peterson 1997):
\begin{equation}
 \tau = -\ln \left( \frac{R_{b}-1+C_{f}}{C_{f}} \right),
\label{eqn:3}
\end{equation}
\begin{equation}
 \log N = \log \left( \frac{\tau b}{f\lambda_{0}}\right) +14.576,
\label{eqn:4}
\end{equation}
where $f$ and $\lambda_{0}$ are oscillator strength and the resonance
wavelength of the absorption line. For some \civ\ doublets, blue
components reach to almost zero flux, and the residual flux is very
uncertain. In that case, we assume that $C_{f} = 1$, and evaluate
column densities by directly fitting the Voigt profile for them.  The
$C_{f}$, $\tau$, and $\log N$ of 30 \civ\ components are listed in
Table 1. For \civ\ lines with $C_{f} < 1$, we also evaluate $\log N$
in the case of $C_{f}=1$ for comparison.

For single lines such as \alii, \siii, and \feii, we evaluated only
the lower limits for $C_{f}$, and $\tau$ by the eqn. (5) and (6),
\begin{equation}
 C_{f} \geq 1 - R_{0},
\label{eqn:5}
\end{equation}
\begin{equation}
 \tau \geq -\ln R_{0}.
 \label{eqn:6}
\end{equation}
We also evaluated $\log N$ in the case of $C_{f}=1$ by the eqn (4).

\alii\ (\zabs\ = 2.4785) and \siii\ (\zabs\ = 2.4785) components in
System B reach to almost zero flux at the line centers. We assume that
$C_{f}=1$ for them, and evaluated column densities by directly fitting
the Voigt profile.

For System A, we evaluated these parameters for only two \civ\ lines
(\zabs\ = 2.4186 and 2.4412), because other lines are heavily blended
with each other, or with the \siii\ lines of System B. For the \civ\
line at \zabs=2.4186, we also evaluated 1 $\sigma$ errors of the
parameters measured in the bottom of the line with 10 \kms\ velocity
width, because this line has very broad line profile (FWHM $> 150$
\kms), and the line center is uncertain.

The \civ\ line (\zabs\ = 2.4186) in System A is likely to be a QIAL,
because the line has $C_{f}=0.31$. On the other hand, the \civ\ line
at \zabs\ = 2.4412 has $C_{f} = 0.8$, although the value has
uncertainty because of its blending with broad \civ\ lines at the same
wavelength region.  If those broad \civ\ lines are removed properly
from the spectrum, the $C_{f}$ of the \civ\ line tends to increase
toward 1.

The result of the covering factor analysis suggests that System B, C
and D cover the quasar completely. The mean $C_{f}$ values of 18, 1,
and 4 \civ\ components in System B, C and D are almost unity, 0.943,
1.00 and 0.945 respectively, and  56\%, 100\%, and 50\% of \civ\
components in these systems have $C_{f}$=1.  Only the \civ\ line at
$z_{abs}$=2.4803 in System B has a exceptionally small covering
factor, $C_{f}$=0.277. But this value is tentative, because this line
is heavily blended with the blue wing of the \civ\ line at
$z_{abs}$=2.4805.

\subsection{Time variability}

Time variability can be a powerful tool in identifying QIALs, simply
because once the variability is confirmed, the line is most likely to
be a QIAL (but QIALs are not always time variable). There were
examples that show this method is indeed very valuable. Absorption
lines with time variability were found at $z_{abs}$=2.13 on the
spectrum of UM675 (Hamann et al. 1995), and at  $z_{abs}$ =2.24 on the
spectrum of Q2343+125 (Hamann et al. 1997b). Equivalent widths of
these lines varied by a factor of 3 and 4 within 12 and 0.3 years,
respectively.

The \feii\ $\lambda 1608$ line at $z_{abs}=2.4367$ in the System A
shows large time variability, though the identification of this line
by D99 probably was not very secure. This line was disappeared in our
spectrum, despite that it had been strong ($W_{rest} = 0.11$ \AA) in
D99 (it is seen around 5525 \AA\ in the lower window of Figure 3).
The quantitative comparison of the equivalent width of the line in our
spectrum with that of D99 is not given, because the spectral
resolution was not identical. Nonetheless, if the line identified by
D99 would be confirmed to be a real \feii\ line at $z_{abs}=2.4367$,
the presence of the time variability is evident and this line is
probably associated with the quasar.

\section{Results and discussion}

Here we classify the detected \civ\ lines into QIALs and SIALs based
on the results in the previous section. After that, we consider the
origin(s) of the absorption lines and the cause of the large number
density of \civ\ lines in the spectrum of high-redshift luminous
quasar HS1603+3820.

\subsection{Result of classification based on three criteria}

\subsubsection{System A}

In System A, all of five broad \civ\ lines have similar smooth line
profiles and they are detected within 3000 \kms\ from each other. One
of them at \zabs\ = 2.4186 which corresponds to $\Delta v$ = $-10600$
\kms\ from HS1603+3820 is confirmed to have a small covering factor of
$C_{f} \sim$ 0.31. For this \civ\ line, FWHM, column density, and
covering factor can be properly measured; they are 190 \kms, 15.8, and
0.31, respectively, and in similar range of values with those of the
already confirmed QIALs, 440 \kms, 15.2 and 0.19 for a \civ\ line at
$-$24000 \kms\ from Q2343+125 (Hamann et al. 1997b), and 56 \kms, 14.1
and 0.31 for a \civ\ line at $-$1500 \kms\ from UM675 (Hamann et
al. 1997a). The other broad lines in System A can not be fitted
individually because of line blending. Nonetheless, they are very
likely to be QIALs not only because they have lower ejection
velocities (nearer to the quasar) than the \civ\ line at the lowest
redshift (\zabs\ = 2.4186) but also because they have similar line
profiles to the fitted \civ\ line.

On the other hand, there is at least one narrow \civ\ line with $b
\sim 16$ \kms\ (\zabs\ = 2.4412) in the system, and the line is
confirmed to have a large covering factor, $C_{f}$=0.8, with the
uncertainty by merging with other broad \civ\ lines. Covering factor
of the line is large, $C_{f}$ is close to unity, which also supports
the idea that the line is a SIAL. If this is true, the association of
the narrow line and the broad lines in System A is just a chance
projection.

There are at least two possible origins for absorbers that produce
QIALs in System A. First, we consider the possibility that they are in
fact similar absorbers that produce much broader lines in the spectra
of the so-called broad absorption line (BAL) quasars. BAL absorbers,
which produces lines with $b$ = 10,000 -- 20,000 \kms\ are thought to
be at the distance of a few tens of parsec from the continuum source
of quasars (cf. Weymann, Turnshek, \& Christiansen 1985; Turnshek et
al. 1985). Although widths of the \civ\ lines in System A with $b$ =
60 -- 120 \kms\ are much smaller than typical BALs, they can be seen
in the spectrum of the quasar at a phase in its transition from (to) a
quasi-BAL to (from) a standard quasar (Morris et al. 1986; Richards et
al. 2002a). When BAL is formed, many narrow components could blend
with each other and make broad absorption feature (BAL) as time
passes. Or when BAL dies, most of the optically thin gas of BAL
absorbers dissipate and only dense cores that produce narrow lines are
left. Ganguly, Charlton, \& Bond (2001) also offered the sporadic (or
quasi-periodic) mass ejection scenario that would have a density
structure of the outflowing wind and produce narrow absorption line
clustering.

Another candidate is outflowing clouds accelerated with the jets from
the central engine of the quasar. In this case, the absorbing gas
extends to distances of hundreds of kpcs (Bridle \& Perley
1984). According to the small covering factor of the \civ\ line at
\zabs\ = 2.4186, however, this interpretation is not favorable because
it is difficult for the absorbers at the distance of hundreds of  kpcs
from the quasar to cover the continuum source partially.

Based on these points, the broad \civ\ lines in System A are probably
produced by outflowing absorbers that are very close to the continuum
source. On the other hand, the narrow \civ\ line at \zabs\ = 2.4412
could be a SIAL which is arisen in an intervening absorber, or a QIAL
which is produced by gas cloud outflowing with almost the same
velocity as absorbers of the broad lines, though they are at high
altitude because its covering factor is close to unity.

Another remarkable point is that \feii\ line at \zabs=2.4367 in D99
shows the time variability, which suggests the line is a
QIAL. However, the relationship between this line and the
corresponding \civ\ line at nearly same redshift is
unclear. Identification of the line as \feii\ $\lambda 1608$ in D99
seems to be based on the detection of \civ\ lines at the same
redshift. In our spectrum in which the line is not seen, however, we
still have the corresponding \civ\ line at nearly the same redshift
(\zabs\ = 2.4366). In fact, the line has relatively narrow line width,
$b = 25.3$ \kms , which can be interpreted as a SIAL.  If this \civ\
line is a really SIAL, we argue that the identification of the line at
5525 \AA\ as \feii\ $\lambda 1608$ in D99 may be reconsidered while
there are other possibilities that (i) there was another \civ\ system
at the same redshift that was disappeared with \feii\, or (ii) the
\civ\ line is a QIAL but only \feii\ line has been disappeared.

\subsubsection{System B}

All of the \civ\ components in System B have narrow line profiles, and
they have large covering factors. System B is plausibly produced by
clumpy gas clouds in an intervening galaxy. The narrow components
distribute within $\pm$ 250 \kms, which can be explained by the bulk
motions in the galaxy. The column density distributions of metal
lines, \civ, \alii, \siii, and \feii, are plotted in Figure
8. High-ionization ion (\civ) distributes more or less evenly within
$\pm$ 250 \kms\ from the center of System B. On the other hand, lines
of low-ionization ions (\alii, \siii, and \feii) with large column
density are seen only near the velocity center; this core-halo
structure of low and high ionization lines also supports that the
lines are arisen in a galaxy. Gas density in halo clouds is so small
that they are easily affected and highly ionized by background UV
flux.

\subsubsection{System C and D}

System C and D are detected within 1000 \kms\ from the quasar
redshift. On one hand, the covering factors of the lines in these
systems are near unity, which suggests that they are separated from
the continuum source far enough to cover the quasar entirely in the
direction of line of sight. On the other hand, their small velocity
differences of 430 \kms\ blueward (System C) and 950 \kms\ redward
(System D) from the quasar redshift imply that they could be
physically associated to the quasar.

There are at least two interpretations for these systems. First, the
lines may be produced by the intervening galaxies in the vicinity of
the quasar that moves around the quasar like cluster (group) of
galaxies. The velocity differences of the systems from the quasar
($\Delta v = 430$ and $950$ \kms) are within the typical velocity
dispersion in rich clusters ($\sigma\sim$1000 \kms, for nearby richest
cluster of galaxies). The large covering factors, $C_{f} \sim 1$, of
these systems can be easily produced by such intervening galaxies.
Secondary, the lines could be produced by gas clouds intrinsically
associated with the quasar, because System C and D do not have
low-ionization lines such as \siii, \alii, and \feii, which suggests
they are on high ionized level. The strong UV flux from the quasar can
make them highly ionized. In the disk-wind model (Murray et al. 1995,
1998) that has been used to explain BAL quasars, intrinsic gas clouds
exist along the wind streamlines from the continuum source to
relatively large distance. The simulations by Proga, Stone, \& Kallman
(2000) show that once outflowing gas clouds reach a high altitude,
they return toward the center along the outflowing stream, which can
produce not only blueshifted but also redshifted absorption lines
relative to the systemic redshift of the quasar
simultaneously. Therefore System C is outflowing from the quasar,
while System D has already reached at a high altitude and flow into
the central source of the quasar. These systems should be separate far
enough from the quasar to cover the quasar entirely.

\subsection{Origin of \civ\ line clustering}

If Systems C and D are arisen in the galaxies around the quasar,
namely they are SIALs, then five of eight \civ\ lines detected in D99
at 2.42 $<$ \zabs $<$ 2.55 are QIALs, and three is SIALs. The number
density of SIALs per unit redshift is $N(z) = 8.82$ at \zabs\ $\sim$
2.38. On the other hand, if Systems C and D are classified into be QIALs,
the corresponding number density is $N(z) = 2.94$, which is almost
consistent with the expected value, $N(z)=2.45^{+0.60}_{-0.49}$, at
\zabs\ $\sim$ 2.40 (Misawa et al. 2002). Therefore the number density
excess of \civ\ absorption lines at \zabs\ $\sim$ \zem\ in the
spectrum of HS1603+3820 may be caused by the number density excess of
QIALs.

Foltz et al. (1986) found that the number density of \civ\ absorption
line increases near the emission redshift of the radio-loud quasars
with their $\sim$ 1 \AA\ resolution spectra. We evaluated the number
densities of \civ\ absorption lines for the 5 quasars in Foltz et
al.'s sample (Q1256+357, Q1416+067, Q1445+335, Q1634+176, and
Q1756+237) that have strong absorption complexes near the quasars. We
used the range of spectra between $z_{15}$ (the redshift $15000$ \kms\
blueward from \zem) and \zem\ or $z_{high}$ (the redshift at which the
\civ\ absorption line of highest-$z$ is detected beyond \zem) over
which the 4$\sigma$ detection limit corresponds to $W_{rest}=$ 0.2
\AA. The mean value of the \civ\ number density (not a Poisson sample)
is $\overline{N(z)} = 36.2 \pm 10.0$ per redshift at \zabs\ $\sim$
1.65. With the $\sim$ 1.5 \AA\ resolution spectrum of D99, the number
density of \civ\ line for HS1603+3820 is evaluated to be 43.5 between
$z_{15}$ and $z_{high}$ (\zabs\ of System D) over which the 4$\sigma$
detection limit is $W_{rest}=$ 0.15 \AA. The $N(z)$ for HS1603+3820
has a large value about the same as the mean value $N(z)$ for 6
quasars in Foltz et al. (1986), though the detection limit of
equivalent width and spectrum resolution are slightly different. These
number density excess of \civ\ lines at \zabs\ $\sim$ \zem\ could be
produced by QIALs.

\section{Conclusions and future work}

We have analyzed the \civ\ absorption line clustering at \zabs\ $\sim$
\zem\ in the spectrum of HS1603+3820 using Subaru$+$HDS data. By using
three properties of \civ\ lines, specifically, time variability,
covering factor, and absorption line profile, we have confirmed that
among four \civ\ systems one system is almost certainly intrinsic to
the quasar, one system is probably arisen in an intervening galaxy.
The other two systems at \zabs\ $\sim$ \zem\ have properties that
are consistent with either interpretation. If they are classified into
QIALs, the number density of SIALs in the Poisson sample becomes to be
almost consistent with the expected value at the similar redshift far
from quasars. In Figure 9, we summarize the distributions of four
\civ\ systems in the physical distance from the quasar in the cases
that Systems C and D are classified into SIALs and QIALs.

We finally note some future prospects. In the course of this study, we
have not applied the time variability analysis to System B, C, and D,
because the resolution of our spectrum is different from that of
D99. An additional spectrum taken with the same configuration as our
spectrum in a few months to years later might enable us to confirm the
time variability of these systems. Deep imaging observation with
narrow-band filters for detecting Ly$\alpha$ or H$\alpha$ emission
lines of galaxies around the quasar will be also useful to understand
the true origins of System C and D. If many galaxies are detected
around the quasar at the redshift around \zem\ of the quasar, they
support the idea that System C and D are produced by intervening
galaxies in the neighbor of the quasar.

\acknowledgments
We are grateful for all the staffs of the Subaru telescope, which is
operated by the National Astronomical Observatory of Japan. We would
like to thank K. Kawabata, S. Kawanomoto, W. Aoki, and N. Suzuki for
their advice about data reduction. We also thank R.F. Carswell,
J.K. Webb, A.J. Cooke, and M.J. Irwin for their VPFIT software package
available in their web site. We wish to thank the anonymous referee
for the report to improve the clarity of this presentation.

\clearpage

\figcaption[Figure 1]{Spectrum of HS1603+3820 at echelle orders of 17,
20 and 23 before normalization. With order 17 and 23, the continuum
function of order 20 is interpolated (dotted line at the middle
window). }
\figcaption[Figure 2]{Normalized Spectrum of HS1603 smoothed by 3
pixels. The obvious cosmic rays are chopped. The lower spectrum in
each window is the 1 $\sigma$ error level.}
\figcaption[Figure 3]{HDS spectrum with R=45000, compared to MMT
spectrum with R $\sim$ 3000 (D99). Four \civ\ systems are surrounded
by dotted lines at both sides. \feii\ 1608 line around 5525 \AA\ in
MMT spectrum disappears in HDS spectrum.}
\figcaption[Figure 4]{\civ\ components of System A. Horizontal axis
denotes the velocity shift from the quasar and the observed
wavelength. Narrow lines at $-9800$ \kms\ $< \Delta v < -9400$ \kms\
are \siii\ 1526 lines of System B. Thick line in the lower right of
figure denotes 10 times width of the instrumental profile (66.7
\kms).}
\figcaption[Figure 5]{\civ\ components of System B. Horizontal axis is
the velocity shift from the quasar. Vertical dotted lines denote the
positions of \civ\ 1548 components. Thick line in the lower right of
\feii\ window denotes 10 times width of the instrumental profile (66.7
\kms).}
\figcaption[Figure 6]{Same as Figure 5, but for System C.}
\figcaption[Figure 7]{Same as Figure 5, but for System D.}
\figcaption[Figure 8]{Distributions of column density for \civ, \siii,
\alii, and \feii\ lines in System B. Open triangle, filled triangle,
filled circle, and filled star denote \civ, \siii, \alii, and \feii\
lines.}
\figcaption[Figure 9]{Cartoons of the structure of absorbers at \zabs\
$\sim$ \zem\ in the line of sight to the quasar HS1603+3820. (A) and
(B) show the distribution of four systems in the physical distance
from the quasar in the cases of Systems C and D being classified into
SIALs and QIALs. Positions of System C and D are exchangeable. (C)
shows the distribution of the systems in the observed velocity shift
from the quasar.}

\clearpage

\begin{deluxetable}{cccccccccc}
\tabletypesize{\scriptsize}
\tablecaption{List of metal absorption lines \label{t1}}
\tablewidth{0pt}
\tablehead{
\colhead{(1)} & 
\colhead{(2)} & 
\colhead{(3)} &
\colhead{(4)} &
\colhead{(5)} &
\colhead{(6)} & 
\colhead{(7)} & 
\colhead{(8)} &
\colhead{(9)} &
\colhead{(10)} \\
\colhead{} & 
\colhead{} & 
\colhead{} & 
\colhead{} & 
\colhead{} & 
\colhead{} & 
\colhead{} &
\colhead{} &
\colhead{$C_{f}=1$} &
\colhead{$C_{f}<1$} \\
\colhead{Line} & 
\colhead{ID} & 
\colhead{$\lambda_{obs}$} & 
\colhead{$z_{abs}$} & 
\colhead{V} &
\colhead{b} &
\colhead{$C_{f}$} &
\colhead{$\tau^{a}$} &
\colhead{$\log N$} &
\colhead{$\log N$} \\
\colhead{} & 
\colhead{} & 
\colhead{(\AA)} & 
\colhead{} & 
\colhead{(km s$^{-1}$)} & 
\colhead{(km s$^{-1}$)} & 
\colhead{} & 
\colhead{} & 
\colhead{(cm$^{-2}$)} &
\colhead{(cm$^{-2}$)} \\
}
\startdata
\tableline
\multicolumn{10}{c}{System A : $z_{abs}=2.44$} \nl
\tableline
 \civ\ $\lambda$1548  & 1  & 5292.7 & 2.4186 & $-$10634 & 114.28$\pm$ 1.96& 0.31$\pm$0.03$^{g}$ 
                              & 4.32$^{+\infty}_{-2.03}$$^{g}$ & 14.23 & 14.80$^{+\infty}_{-0.28}$$^{g}$ \\
                      & 2  & 5303.8 & 2.4257 & $-$10008 & 77.68$\pm$ 5.59$^{f}$& ... & ... & ... & ... \\
                      & 3  & 5316.7 & 2.4341 & $-$9276  & 65.19$\pm$ 2.85$^{f}$& ... & ... & ... & ... \\
                      & 4  & 5320.0 & 2.4362 & $-$9092  & 80.06$\pm$ 1.52$^{f}$& ... & ... & ... & ... \\
                      & 5  & 5320.6 & 2.4366 & $-$9058  & 25.30$\pm$ 1.16$^{f}$& ... & ... & ... & ... \\
                      & 6  & 5327.7 & 2.4412 & $-$8656  & 15.68$\pm$ 0.31&0.796 & 1.912 & 13.4 & 13.6 \\
                      & 7  & 5342.9 & 2.4510 & $-$7806  & 74.36$\pm$ 6.06$^{f}$& ... & ... & ... & ... \\
\nl
\tableline
\multicolumn{10}{c}{System B : $z_{abs}=2.48$} \nl
\tableline
 \civ\ $\lambda$1548  & 1  & 5381.8 & 2.4762 & $-$5628  &  9.62 $\pm$ 0.27 & 1.000 & 1.169 & 13.2 & ...  \\                  
                      & 2$^{b}$ & 5382.2 & 2.4764 & $-$5608  &  7.39 $\pm$ 0.11 & 1.000 & ... & 14.0$^{d}$ & ...  \\  
                      & 3  & 5382.8 & 2.4768 & $-$5573 & 20.42 $\pm$ 1.92 & 1.000 & 0.607 & 13.2 & ...  \\                  
                      & 4  & 5383.0 & 2.4769 & $-$5560 &  4.62 $\pm$ 0.60 & 1.000 & 0.702 & 12.6 & ...  \\                  
                      & 5  & 5383.3 & 2.4771 & $-$5544 &  7.57 $\pm$ 2.09 & 1.000 & 0.328 & 12.5 & ...  \\                  
                      & 6  & 5383.8 & 2.4774 & $-$5519 & 13.49 $\pm$ 4.94 & 0.977 & 5.653 & 13.8 & 14.0 \\                  
                      & 7$^{b}$ & 5383.9 & 2.4775 & $-$5512 &  7.76 $\pm$ 0.54 & 1.000 & ... & 14.1$^{d}$ & ...  \\ 
                      & 8  & 5384.2 & 2.4777 & $-$5494 &  8.84 $\pm$ 2.62 & 0.783 & 0.958 & 12.9 & 13.1 \\                  
                      & 9  & 5384.5 & 2.4779 & $-$5477 &  9.63 $\pm$ 0.53 & 0.985 & 2.416 & 13.5 & 13.5 \\                  
                      & 10 & 5385.0 & 2.4782 & $-$5450 & 12.37 $\pm$ 0.44 & 1.000 & 3.258 & 13.7 & ...  \\                  
                      & 11 & 5385.6 & 2.4786 & $-$5416 & 24.03 $\pm$ 2.29 & 0.997 & 3.128 & 14.0 & 14.0 \\       
                      & 12 & 5386.2 & 2.4790 & $-$5385 & 10.65 $\pm$ 1.25 & 0.980 & 4.072 & 13.7 & 13.8 \\                  
                      & 13 & 5386.4 & 2.4791 & $-$5373 & 20.78 $\pm$ 3.19 & 0.985 & 1.917 & 13.7 & 13.7 \\                  
                      & 14 & 5387.1 & 2.4795 & $-$5336 &  8.48 $\pm$ 0.81 & 1.000 & 0.645 & 12.9 & ...  \\                  
                      & 15 & 5387.4 & 2.4797 & $-$5319 &  8.57 $\pm$ 0.24 & 1.000 & 2.870 & 13.5 & ...  \\       
                      & 16 & 5388.3 & 2.4803 & $-$5268 & 23.13 $\pm$ 1.43 & 0.277$^{e}$& 0.901 & 12.8 & 13.5 \\                  
                      & 17 & 5388.6 & 2.4805 & $-$5251 &  5.47 $\pm$ 0.12 & 0.987 & 5.931 & 13.5 & 13.6 \\       
                      & 18 & 5389.8 & 2.4813 & $-$5182 & 14.49 $\pm$ 2.02 & 1.000 & 0.131 & 12.4 & ...  \\
 \alii\ $\lambda$1670 & 1  & 5811.2 & 2.4781 & $-$5458 & 12.09 $\pm$ 0.65 & $>$0.336 & $>$0.410  & 11.8 & ... \\                   
                      & 2$^{c}$ & 5811.9 & 2.4785 & $-$5422 &  9.69 $\pm$ 0.16 & 1.000 & ... & 12.9$^{d}$ & ... \\ 
                      & 3  & 5813.4 & 2.4794 & $-$5345 &  5.25 $\pm$ 1.13 & $>$0.140 & $>$0.151  & 11.0 & ... \\                   
                      & 4  & 5813.9 & 2.4797 & $-$5322 &  6.37 $\pm$ 0.44 & $>$0.361 & $>$0.448  & 11.6 & ... \\  
                      & 5  & 5815.2 & 2.4805 & $-$5251 &  4.00 $\pm$ 0.19 & $>$0.583 & $>$0.875  & 11.7 & ... \\  
                      & 6  & 5816.4 & 2.4812 & $-$5193 &  4.54 $\pm$ 0.45 & $>$0.373 & $>$0.466  & 11.4 & ... \\                   
                      & 7  & 5816.6 & 2.4813 & $-$5181 &  7.15 $\pm$ 1.34 & $>$0.225 & $>$0.255  & 11.4 & ... \\                   
 \siii\ $\lambda$1526 & 1  & 5306.3 & 2.4757 & $-$5670 & 11.46 $\pm$ 0.35 & $>$0.727 & $>$1.298  & 13.5 & ... \\                 
                      & 2  & 5307.5 & 2.4764 & $-$5606 & 12.22 $\pm$ 1.40 & $>$0.296 & $>$0.351  & 13.0 & ... \\
                      & 3  & 5310.1 & 2.4781 & $-$5459 & 11.23 $\pm$ 0.51 & $>$0.322 & $>$0.388  & 13.0 & ... \\                 
                      & 4$^{c}$ & 5310.7 & 2.4785 & $-$5422 & 10.73 $\pm$ 0.14 & 1.000 & ... & 14.1$^{d}$ & ... \\
                      & 5  & 5312.5 & 2.4797 & $-$5322 &  8.15 $\pm$ 0.37 & $>$0.398 & $>$0.507  & 12.9 & ... \\
                      & 6  & 5313.7 & 2.4805 & $-$5252 &  4.93 $\pm$ 0.11 & $>$0.803 & $>$1.625  & 13.2 & ... \\
                      & 7  & 5314.8 & 2.4812 & $-$5193 &  5.63 $\pm$ 0.31 & $>$0.497 & $>$0.687  & 12.9 & ... \\                 
                      & 8  & 5315.1 & 2.4814 & $-$5174 & 13.30 $\pm$ 0.55 & $>$0.464 & $>$0.623  & 13.2 & ... \\
 \feii\ $\lambda$1608 & 1  & 5595.1 & 2.4786 & $-$5422 &  5.59 $\pm$ 0.27 & $>$0.580 & $>$0.868  & 13.3 & ... \\
\nl
\tableline
\multicolumn{10}{c}{System C : $z_{abs}=2.54$} \nl
\tableline
 \civ\ $\lambda$1548  & 1  & 5475.9 & 2.5369 & $-$431 & 11.11 $\pm$ 0.19 & 1.000 & 0.860 & 13.1 & ...  \\
\nl
\tableline
\multicolumn{10}{c}{System D : $z_{abs}=2.55$} \nl
\tableline
 \civ\ $\lambda$1548  & 1  & 5500.2 & 2.5526 & $+$896 &  9.40 $\pm$ 0.51 & 0.787 & 0.864 & 12.9 & 13.0 \\
                      & 2  & 5500.5 & 2.5528 & $+$914 &  7.64 $\pm$ 1.49 & 1.000 & 0.297 & 12.5 & ...  \\
                      & 3  & 5500.9 & 2.5531 & $+$939 & 12.48 $\pm$ 0.30 & 0.991 & 6.813 & 13.9 & 14.0 \\
                      & 4$^{b}$ & 5501.5 & 2.5535 & $+$970 & 10.14 $\pm$ 0.15 & 1.000 & ... & 14.6$^{d}$ & ... \\
\tablenotetext{a}{Optical depth: For \civ\ line, it is evaluated for
bluer component in the case $C_{f}$ is the value of column (7). For
single line, it is evaluated in the case of $C_{f}$=1.0.}
\tablenotetext{b}{Center of blue component reaches to almost zero
flux.}
\tablenotetext{c}{Center of line reaches to almost zero flux.}
\tablenotetext{d}{It is evaluated by fitting the Voigt profile
directly.}
\tablenotetext{e}{This line is blended with the blue wing of \civ\
line at \zabs=2.4805.}
\tablenotetext{f}{It is evaluated for blue component.}
\tablenotetext{g}{One sigma error is evaluated with 10 \kms\ velocity
width, because line profile is very broad and line center is uncertain.}
\enddata
\end{deluxetable}

\clearpage
\begin{figure}
  \plotone{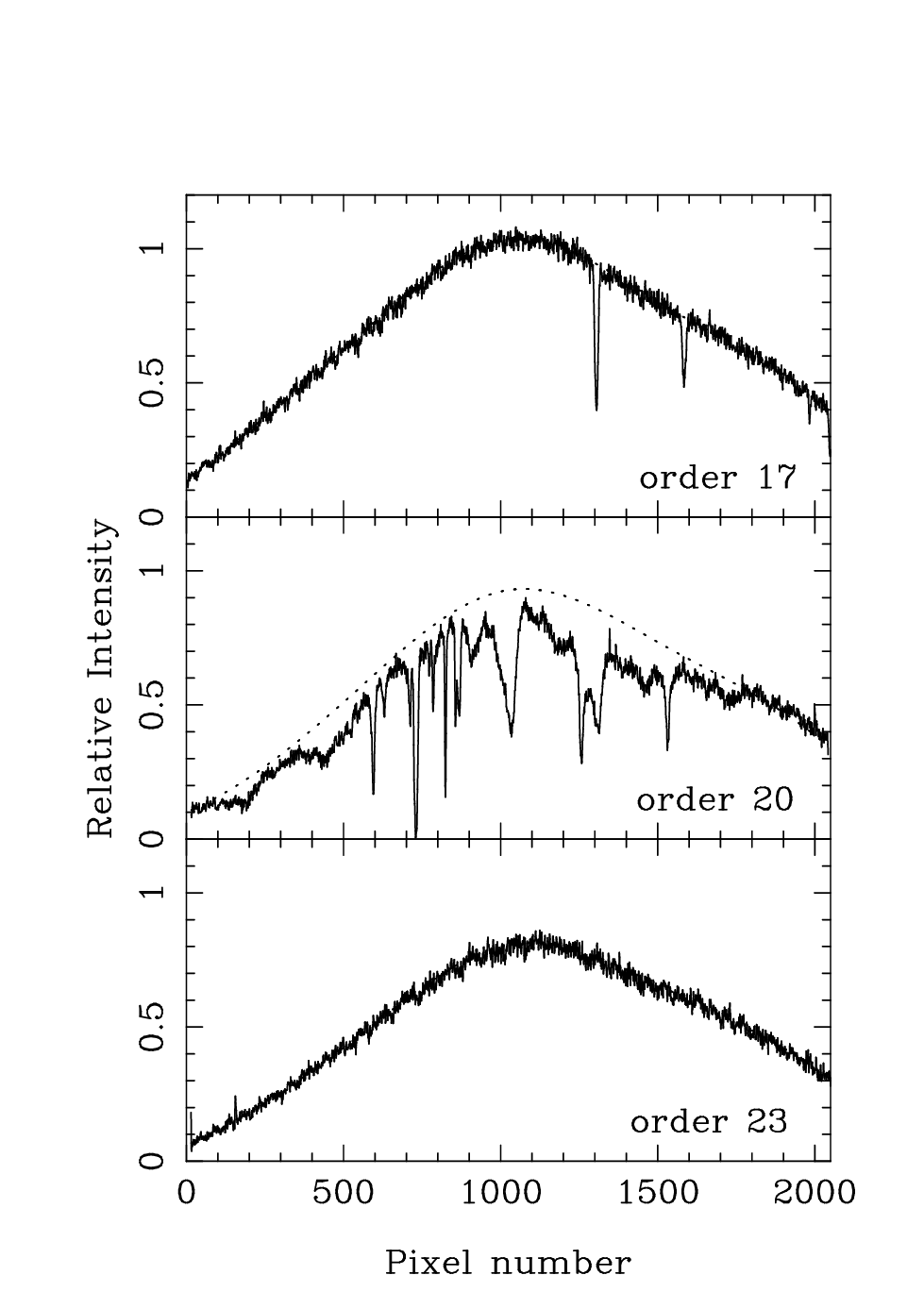}
\\Fig.~1
\end{figure}
\clearpage

\clearpage
\begin{figure}
  \plotone{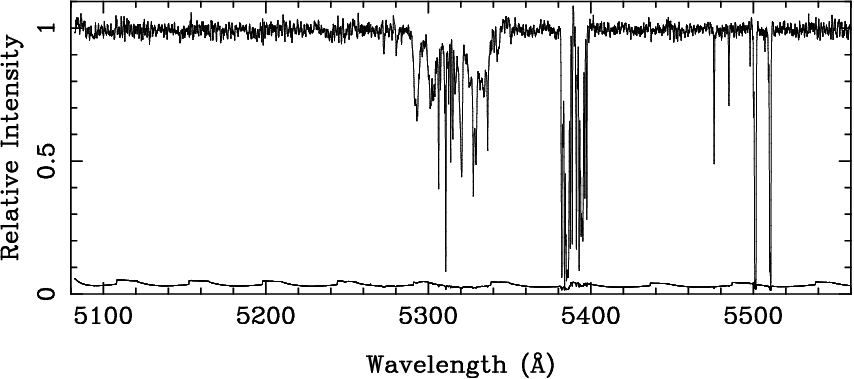}
  \plotone{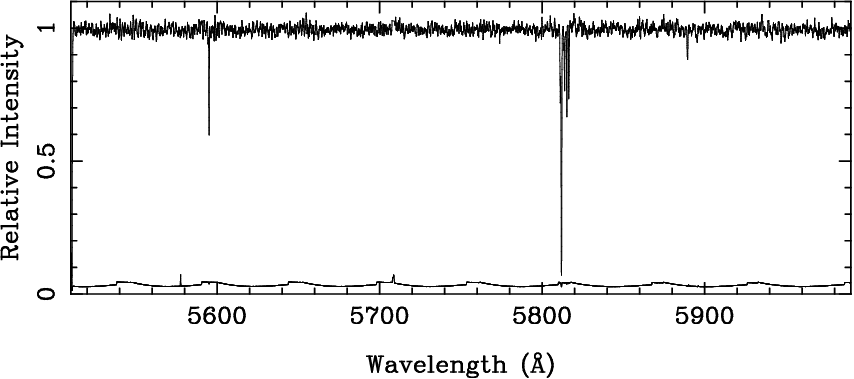}
  \plotone{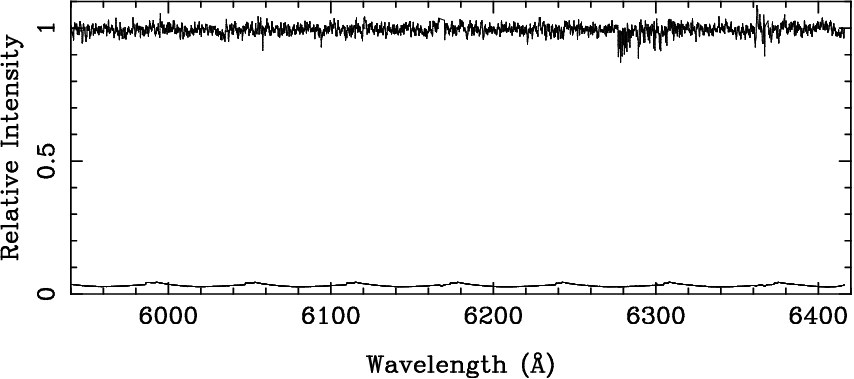}
\\Fig.~2
\end{figure}
\clearpage

\clearpage
\begin{figure}
  \plotone{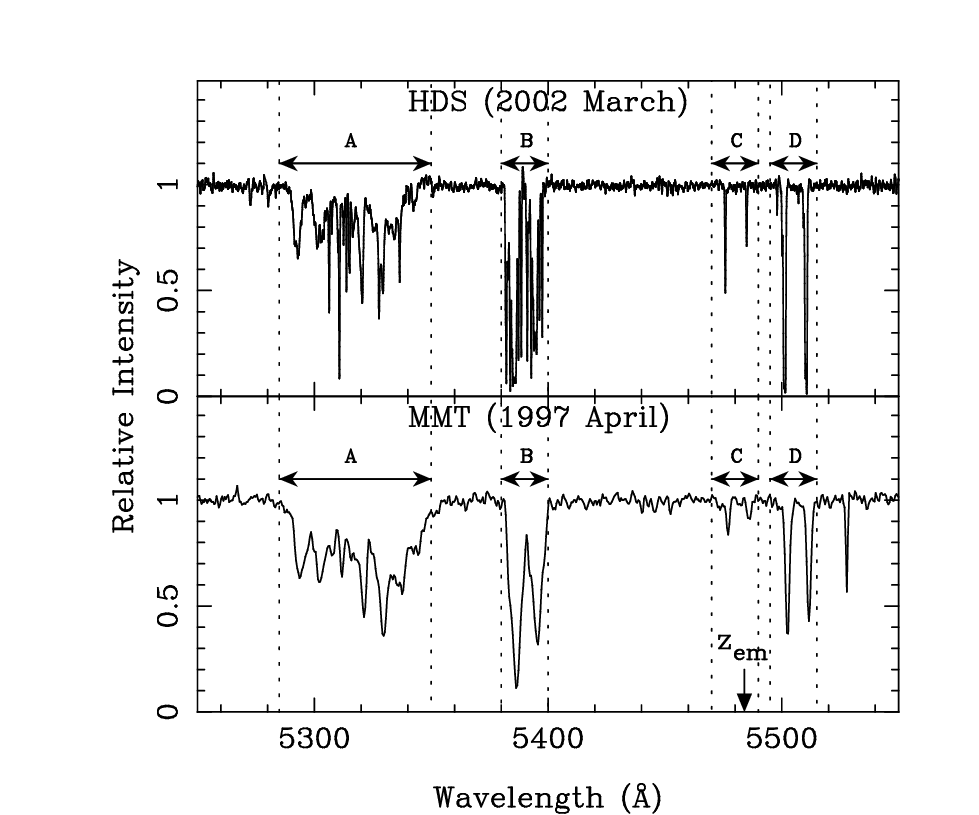}
\\Fig.~3
\end{figure}
\clearpage

\clearpage
\begin{figure}
  \plotone{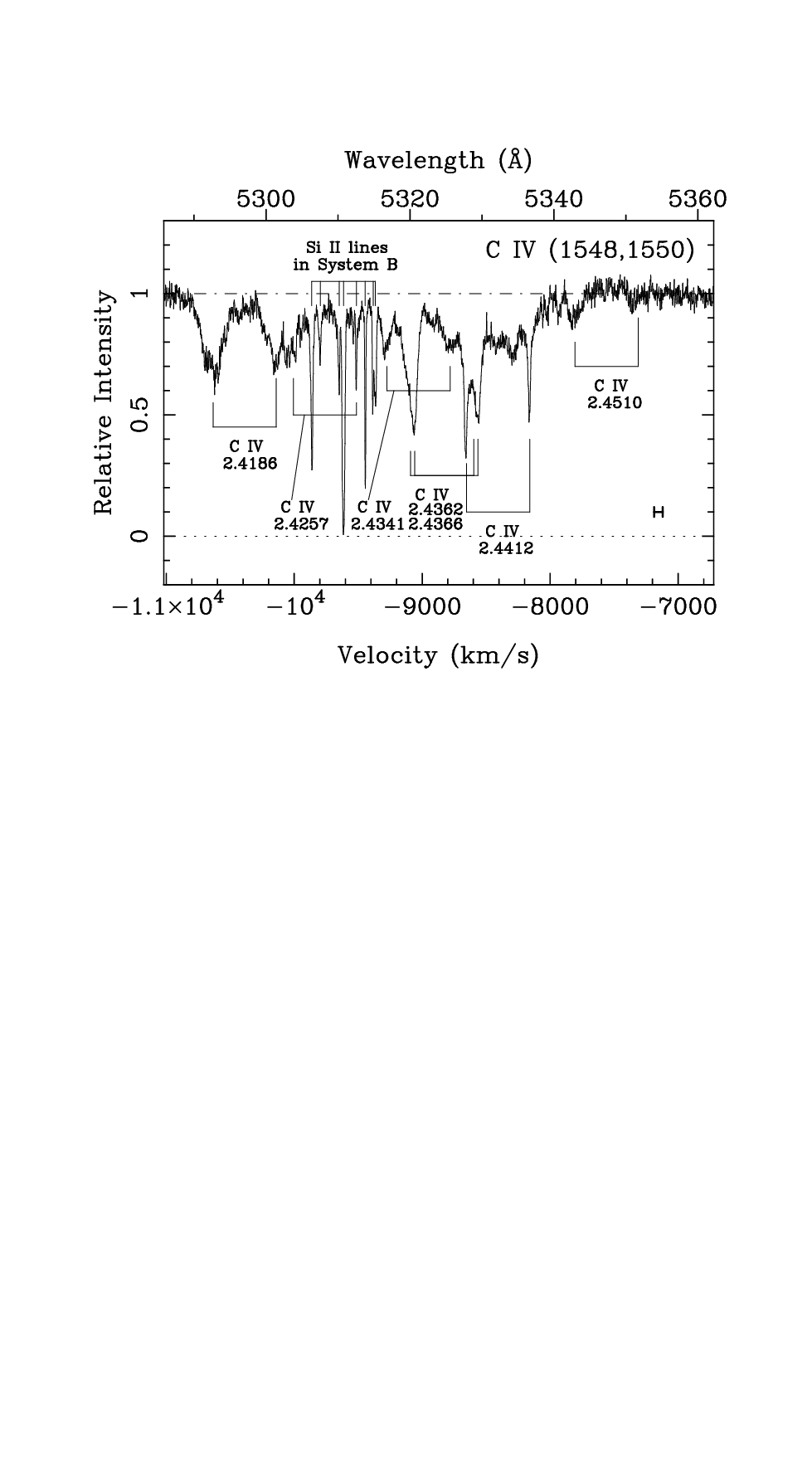}
\\Fig.~4
\end{figure}
\clearpage

\clearpage
\begin{figure}
  \plotone{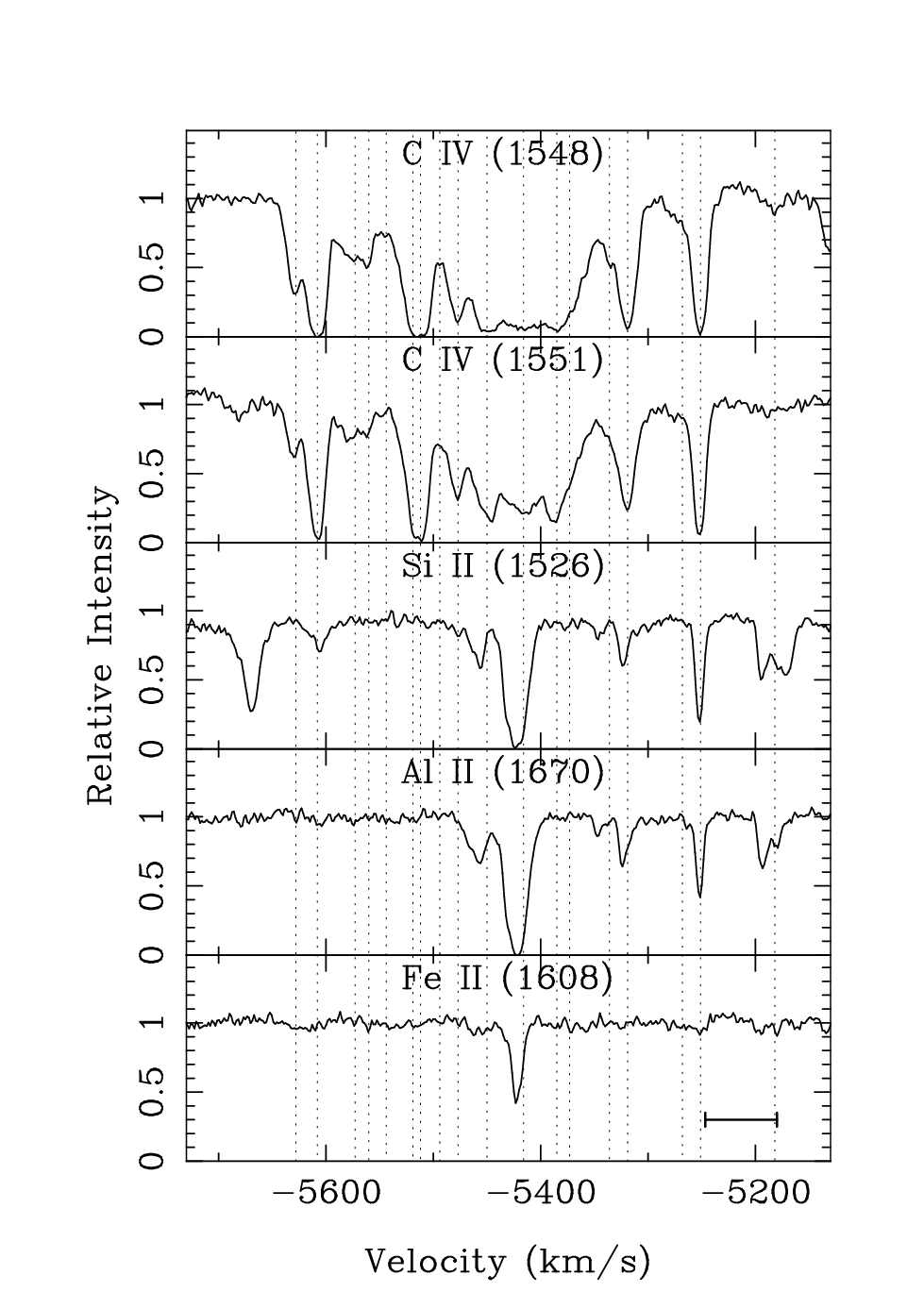}
\\Fig.~5
\end{figure}
\clearpage

\clearpage
\begin{figure}
  \plotone{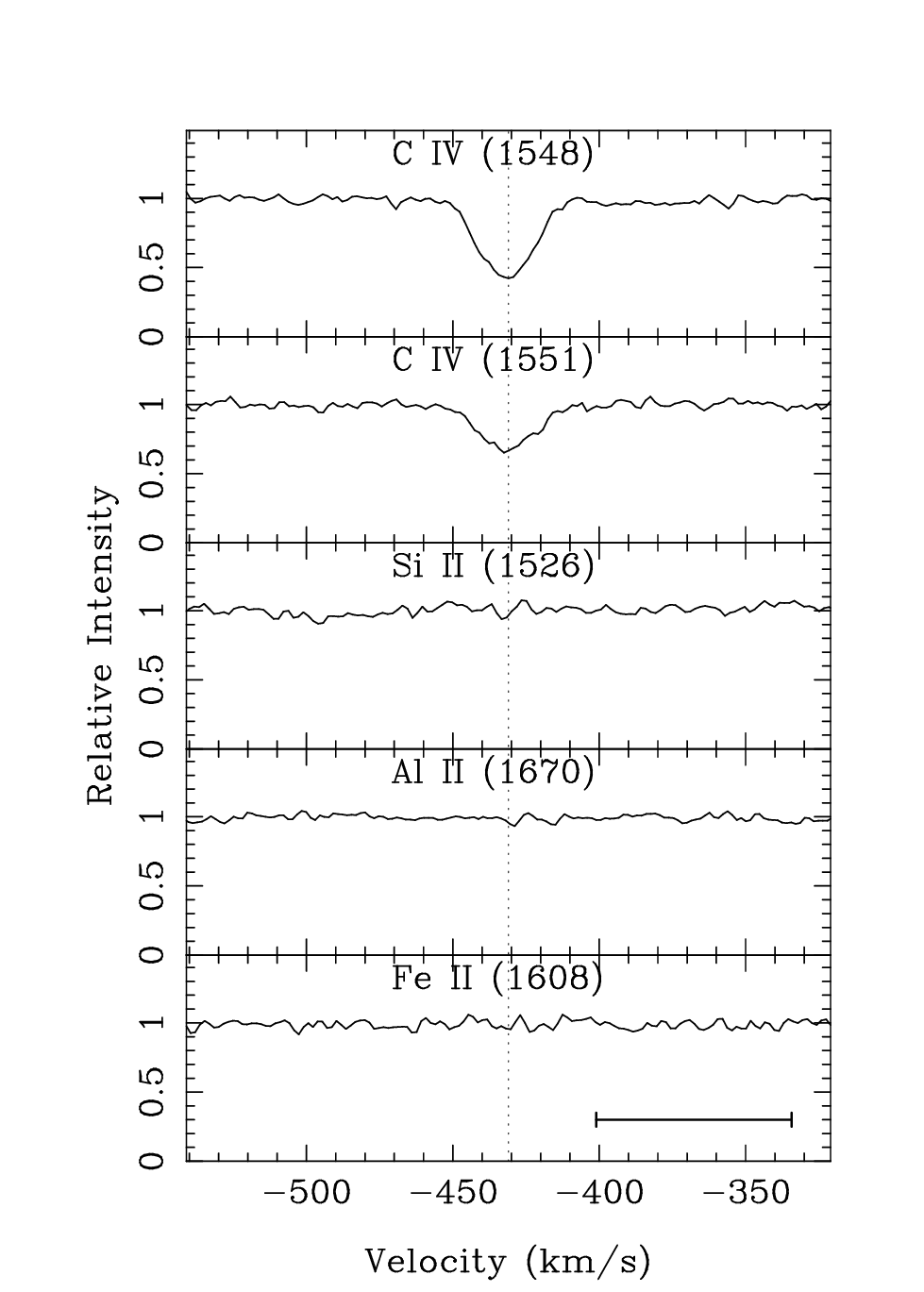}
\\Fig.~6
\end{figure}
\clearpage

\clearpage
\begin{figure}
  \plotone{misawa.fig7.ps}
\\Fig.~7
\end{figure}
\clearpage

\clearpage
\begin{figure}
  \plotone{misawa.fig8.ps}
\\Fig.~8
\end{figure}
\clearpage

\clearpage
\begin{figure}
  \plotone{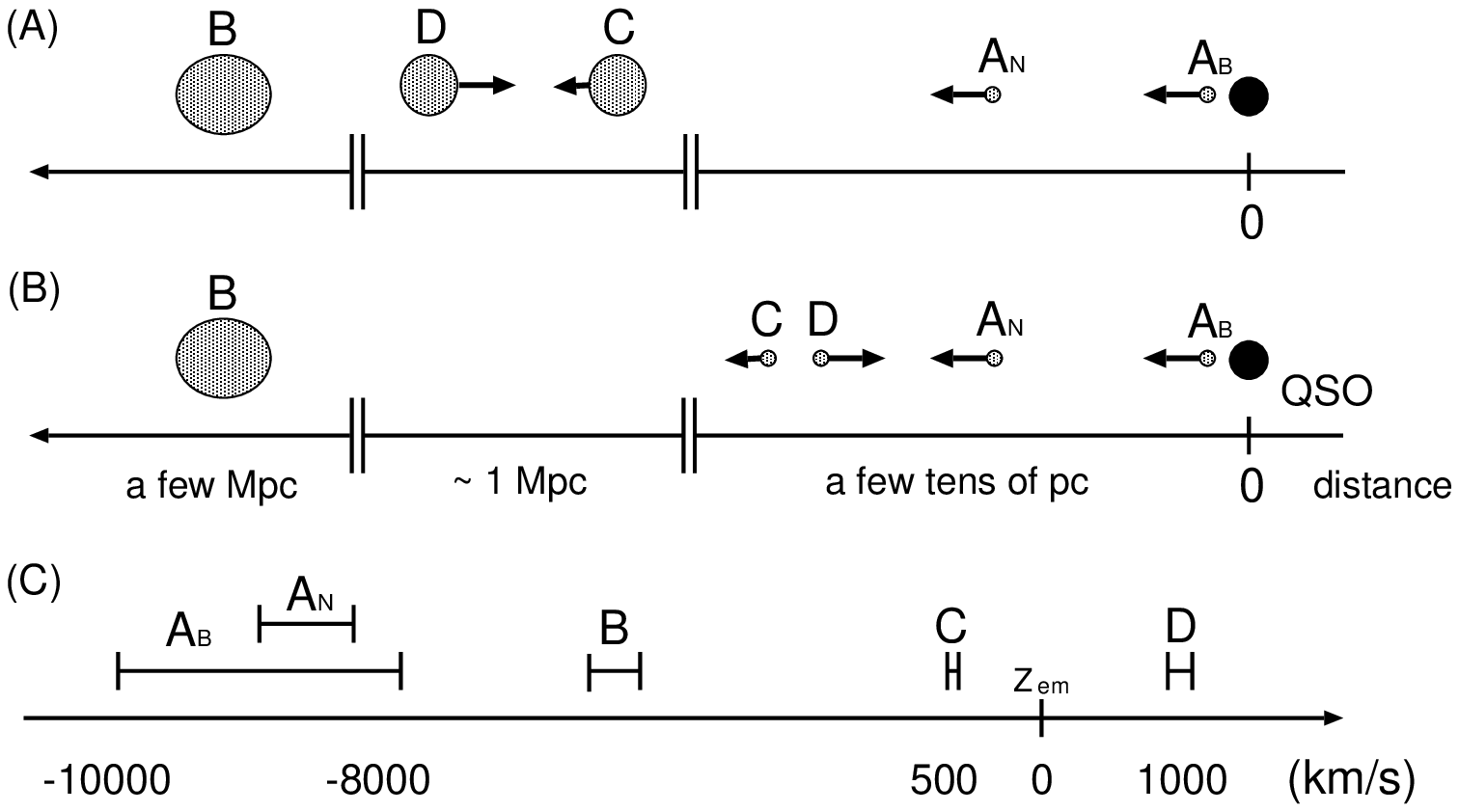}
\\Fig.~9
\end{figure}
\clearpage

\end{document}